# BOND RUPTURE MECHANISM ENABLES TO EXPLAIN IN BLOCK ASYMMETRY OF RELAXATION, FORCE-VELOCITY CURVE AND THE PATH OF ENERGY DISSIPATION IN MUSCLE

E.V. Rosenfeld

*I.     Inttroduction*

В середине пятидесятых годов прошлого века были достигнуты огромные успехи в понимании механизма мышечного сокращения, см. Huxley A F (2000). Было установлено (Huxley H E 1953; Huxley H E & Hanson 1954; Huxley A F & Niedergerke 1954; Huxley H E (1957)) that shortening of the muscle fibre takes place by relative sliding movement of two sets of filaments – actin (AF) and myosin (MF) lines. Выяснилось, что они приводятся в движение за счет циклической конформации myosin heads (MHs), которые последовательно attach to AF forming a cross-bridge, совершают гребковое движение (hoe-type) и снова отделяются. В каждом цикле мостик движется и совершает механическую работу за счет энергии, получаемой при гидролизе одной молекулы ATP, и КПД of the chemo-mechanical transduction зависит от скорости сокращения мышцы.

Первая феноменологическая теория, построенная на основе этих данных, была создана by Huxley A F (1957). В ней предполагалось, что в начале каждого цикла MH получает в результате гидролиза ATP запас энергии $E_{ATP}$, которая затем по мере непрерывной конформации в процессе working stroke постепенно расходуется. За счет подбора зависимостей от координаты мостика attachment and detachment probabilities and величины и знака генерируемой им силы эта теория позволяла вполне успешно моделировать основные особенности поведения мышцы. Однако, уже через несколько лет выяснилось (Podolsky 1960; Huxley and Simmons 1971), что при скачкообразном изменении длины или скорости сокращения мышцы возникает странная асимметрия. Оказалось, что на начальной стадии релаксация после резкого удлинения происходит значительно медленнее, чем после резкого укорочения. Для того, чтобы объяснить это явление Huxley and Simmons (1971) создали новую модель, в которой предполагалось, что у MH есть лишь несколько дискретных состояний, между которыми она совершает быстрые переходы, а (медленная) конформация в процессе working stroke связана с постепенным изменением вероятностей заполнения этих состояний.

Обе модели (Huxley A F 1957) и (Huxley and Simmons 1971) являются чисто феноменологическими и оставляют без ответа многие принципиально важные вопросы, касающиеся механизма работы of myosin motor (MM). Тем не менее, уже сам основополагающий принцип модели – представление о том, является ли конформация мостика в процессе working stroke непрерывной или дискретной – практически однозначно предопределяет ответ на важнейший вопрос о том, какие силы вызывают эту конформацию. С одной стороны, существует множество работ, так или иначе связанных с моделью (Huxley and Simmons 1971). В них обосновывается идея о том, что ММ приводится в движение by Brownian ratchet or some other kind of Maxwell's demon, able to extract useful work directly from random fluctuations (heat).

С другой стороны, вполне в духе модели (Huxley A F 1957), представляется очевидным, что MH conformation может происходить только под действием внутренних сил, которые возникают и совершают работу в процессе гидролиза ATP. Поскольку только электромагнитные взаимодействия определяют ход любых химических процессов, можно утверждать, что direct conversion of chemical energy to mechanical one is utilization of the mechanical work performed by the Coulomb forces when rearranging the electrical charges in the course of chemical reaction. Если согласиться с этим утверждением, то для выяснения механизма работы ММ необходимо исследовать перераспределение зарядов, которое в процессе гидролиза и конформации происходит внутри MH (см. в этой связи Rosenfeld 2012). В противном случае либо ММ приводится в действие неким взаимодействием, отличным от электромагнитного, либо внутри MH следует искать структуры, выполняющие функции демона Максвелла.

## II. Irreversibility and asymmetry of relaxation

Еще один вопрос исключительной важности связан с причиной возникновения асимметрии релаксационных процессов в мышце. В классической молекулярно-кинетической теории (see e.g, Feynman et al. 1963) необратимость в молекулярных масштабах отсутствует, и любая попытка объяснить эту асимметрию процессами, происходящими внутри MH, снова ведет к демону Максвелла. Именно поэтому Huxley and Simmons (1971) пришлось ввести крайне экзотическую зависимость энергии от координаты (see Fig.6 therein), на которой последовательные все более высокие ступени как раз и описывают ratchet gear.

Ситуация представляется гораздо менее запутанной если полагать, что эта асимметрия связана с процессами, происходящими вне MHs. Стоит лишь предположить, что в процессе сокращения и растяжения саркомера на филаменты действует обычная сила трения, препятствующая их скольжению, и асимметрия возникает сама собой. Теперь при резком растяжении саркомера действует сила трения скольжения, которая после завершения twitch сменяется силой трения покоя. В обоих случаях сила трения и упругие силы, возникающие внутри MHs, действуют на AF в одном направлении и складываются. Если теперь сила трения покоя уменьшается со временем, возникают компенсирующие это изменение дополнительные упругие растяжения внутри MHs. Поэтому общая сила, действующая на AF, будет уменьшаться только по мере развития конформационных процессов.

При резком сокращении длины саркомера силы, вызывающие скольжение филаментов на начальном этапе – это упругие силы внутри MHs. Теперь сила трения и упругие силы направлены в разные стороны и вычитаются. Если сила трения исчезает, все упругие силы оказываются приложенными непосредственно к Z-line, так что тензодатчик будет фиксировать возрастание muscle fibre tension по мере уменьшения силы трения. Следовательно, причиной асимметрии релаксационных процессов может быть исчезающая за время порядка 0.1 ms сила трения между филаментами (в связи с этим см. также (Rosenfeld 2014)).

Таким образом, «центр тяжести» проблемы асимметрии релаксации смещается, и на первый план выходит еще один принципиально важный вопрос, ответ на который мы пытаемся дать в этой работе. Хорошо известно (см., например, Appendix E in Ford, Huxley and Simmons (1977)) что при характерных для мышцы скоростях скольжения филаментов (at most few μm s$^{-1}$) обычные гидродинамические силы трения и совершаемая ими работа пренебрежимо малы и не способны обеспечить диссипацию значительной (порядка $10\, k_B T$ per MH) механической энергии. Поэтому остается не известным канал диссипации той части энергии $E_{ATP}$, которая не была превращена в механическую работу.

Не исключено, что работа (Bormuth et al. 2009) дает ключ к решению этой проблемы. Using optical tweezers, the authors of the work measured the frictional drag force of individual kinesin-8 motor proteins interacting with their microtubule tracks. They also argue that these frictional forces arise from breaking bonds between the motor domains and the microtubule. Естественно предполагать, что такого же типа hindering force должна действовать на те MHs, которые aren't stereospecifically attached to AF и скользят мимо него в процессе muscle contraction. Часть из них может, видимо, создавать weak bonds, например, прикрепляясь nonstereospecifically to AF, (Bershitsky et al. 1997 ), и тогда потери энергии при скольжении филаментов равны работе, затрачиваемой на разрыв этих связей. Таким образом, мы приходим к bond rupture dynamics, которая, видимо, впервые была использована для описания биологических объектов by Bell (1978). Позднее проблема активно исследовалась в разных аспектах (Tawada K and Sekimoto K (1991); Evans and Ritchie (1997); Suda (2001); Filippov et al. (2004)) вплоть до последнего времени (Bar-Sinai et al. 2015). В разделе III мы рассмотрим простейшие модели, позволяющие количественно описать силу сопротивления, возникающую при скольжении филаментов из-за weak-binding interaction between AF and MHs. В разделе IV обсуждается возможность использования полученных результатов для описания force-velocity curve.

### III. Bond rupture dynamics in muscle

Рассмотрим ансамбль из $N$ штук MHs, расположенных периодически вдоль MF с шагом $d_m$, см. Fig.1. Из них $N_a$ головок are stereospecifically attached to AF и прикладывают к нему силу, вызывающую сокращение мышцы. Остальные $N_d = N - N_a$ MHs are detached from AF, который движется относительно них в направлении оси $\hat{x}$ со скоростью $v$. В общем случае следует предположить, что с шагом $d_a$ вдоль AF расположены active sites шириной $\delta$, внутри которых MHs могут nonstereospecifically attach to AF, создавая weak bonds. Если полагать, что периоды $d_a$ и $d_m$ несоизмеримы, при больших $N_d$ в любой момент $N_d \delta/d_a$ MHs должно находиться напротив этих active sites. Обозначив вероятность возникновения связи МН с AF в этом положении через $w$, а через $U$ - энергию такой связи, т.е. работу, которую нужно затратить, чтобы протащить МН над активным центром, получим:

$$F_h = N_d f_h, \quad f_h = w U/d_a, \tag{1}$$

где через $f_h$ обозначена hindering force в расчете на одну detached МН.

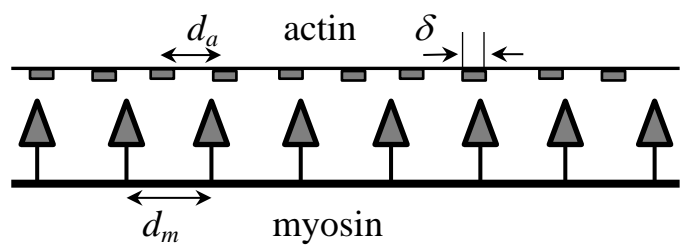

**Figure 1**. Myosin heads периодически прикрепленные к thick filament с шагом $d_m$, могут «прилипать» к thin filament внутри интервалов длиной $\delta$, распределенных периодически с шагом $d_a$.

По сути дела, в этом элементарном подходе мы просто постулировали, что на каждую МН, движущуюся над active site, действует тормозящая сила $U/\delta$. Слегка усложнив задачу, можно предположить, что внутри MH имеется elastic element EE с упругим коэффициентом $k$, и когда AF смещается на $x$, точка закрепления MH на нем не меняется, но EE растягивается на $x$. В результате каждая связь создает hindering force $kx$, так что полная hindering force равна

$$F_h = k \sum_i x_i . \tag{2}$$

Суммирование здесь идет по всем связям, а $x_i$ – внутренняя координата $i$-ой связи, т.е. перемещение филаментов, которое произошло с момента ее возникновения. Чтобы получить возможность написать кинетические уравнения, будем полагать также, что

- среднее время возникновения связи равно $K_+^{-1}$;
- тепловые флуктуации могут разрушить связь, и если действует только этот механизм, ее среднее время жизни $\tau = K_-^{-1}$ не зависит от скорости скольжения филаментов $v$ и силы натяжения EE;
- в любом случае связь разрывается когда ее сила натяжения достигает предельного значения $kx_0$.

Таким образом, теперь смысл введенной величины энергии связи $U$ в том, чтобы определить предельную силу натяжения (скажем, через равенство $kx_0^2 = 2U$), либо время жизни связи (скажем, через равенство $\tau \sim \exp(U/k_B T)$. Хотя эти простейшие модельные предположения вряд ли реалистичны, они позволяют предельно упростить расчеты и облегчить понимание физического смысла рассматриваемого механизма трения.

Предположим сначала, что $x_0 \gg v\tau$, т.е. практически всегда флуктуационный механизм срабатывает раньше, чем связь рвется из-за слишком сильного натяжения. Тогда средняя длина EE в момент разрыва связи равна $v\tau$, а средняя за время ее существования сила, которую она прикладывает к AF, равна $kv\tau/2$. Поскольку только флуктуационные процессы играют роль при образовании и исчезновении связей, их среднее количество пропорционально отношению rate constants $p = K_+/(K_+ + K_-)$ и некому геометрическому фактору $\eta$, зависящему от $d_a$ and $\delta$. Последний для нас интереса не представляет, но важно, что количетсво связей снова пропорционально the number of detached MHs $N_d$:

$$F_h(v) = N_d f_h, \quad f_h = p\eta k v\tau / 2 . \tag{3}$$

Таким образом, в этом случае hindering force прямо пропорциональна скорости скольжения филаментов $v$.

В противоположном предельном случае $x_0 \ll v\tau$ связи достигают максимальной длины и рвутся гораздо раньше, чем их успевают разрушить флуктуации. Поэтому средняя длина связи равна $x_0/2$, а тормозящая сила

$$F_h(v) = N_d f_h, \quad f_h = \frac{kx_0\eta K_+}{2\left(K_+ + v/x_0\right)} \underset{\text{if } \frac{v}{x_0 K_+} \ll 1}{\approx} \frac{kx_0\eta}{2}\left(1 - \frac{v}{x_0 K_+}\right). \quad (4)$$

Здесь снова $\eta$ означает геометрический фактор, а зависимость от скорости можно определить из следующих соображений. В равновесном состоянии с равной вероятностью существуют связи любой длины от 0 до $x_0$, так что за время $dt$ должна разрываться доля $vdt/x_0$ от их полного числа $n$. Возникнуть за то же время должно $(N_d\eta - n)K_+ dt$ связей, и приравнивая эти два числа, мы и получаем (4). Если исключить из рассмотрения геометрический фактор, предположив, что $\delta = d_a$, так что связь может возникать в любой точке, то решив простое кинетическое уравнение, нетрудно получить формулу

$$F_h\left(u = \frac{x_0}{v\tau}\right) = \frac{1}{2}N_d kv\tau \frac{1 - (1+u)\exp(-u)}{K_-/K_+ + 1 - \exp(-u)}, \quad (5)$$

справедливую при любом соотношении между $x_0$ and $v\tau$. При малых $v$ эта функция линейно растет в соответствии с (3), потом достигает максимума и выходит на гиперболу (4). В области левее максимума эта функция ведет себя примерно так же, как экспериментальные кривые в (Bormuth 2009).

Таким образом, в разных предположениях мы получаем разные зависимости hindering force от скорости сокращения мышцы[1]. В следующем разделе мы проанализируем, как функция $F_h(v)$ может быть связана с force-velocity curve.

## IV. Hill's force-velocity curve

В isovelocity contraction state каждая МН проводит некоторое время $t_w$, совершая working stroke, затем в течение времени, которое мы назовем recovery time $t_r$, происходит подготовка к совершению нового working stroke, и цикл повторяется. Пытаясь разобраться, на что именно тратится энергия $E_{ATP}$, которую МН получает при гидролизе ATP, и почему сила, создаваемая мышцей, может зависеть от скорости ее сокращения, рассмотрим еще одну очень простую модель. По сути она сводится к двум предположениям, определяющим зависимости времен $t_w$ and $t_r$ от скорости скольжения филаментов:

- $t_r$ зависит только on chemical reaction rates, но не от $v$.
- В течение working stroke при любой скорости MH совершает одну и ту же работу $E_{ATP}$ при одном и том же сдвиге на величину шага $S \approx 10\,\text{nm}$.

---

[1] Стоит отметить здесь, что падающая с увеличением скорости сила трения при определенных условиях может вызвать нестабильность движения.

Эти предположения вполне соответствуют не только духу, но отчасти и букве модели (Huxley 1957); некоторые дополнительные аргументы в их пользу можно найти в (Rosenfeld 2012). Из них мы получаем: (i) средняя сила, которую во время working stroke прикладывает к AF одна MH, равна

$$f_0 = E_{ATP}/S\,; \tag{6}$$

(ii) продолжительность цикла MH равна

$$T(v) = t_w + t_r = S/v + t_r\,; \tag{7}$$

(iii) доля MHs, совершающих в данный момент working stroke, составляет

$$\frac{t_w}{t_w + t_r} = \left(1 + v/v_0\right)^{-1}, \quad v_0 = S/t_r\,. \tag{8}$$

(iv) propelling force, которую прикладывает к AF ансамбль из $N$ MHs, равна

$$F_{prop}(v) = N f_0 \left(1 + v/v_0\right)^{-1}\,. \tag{9}$$

Связь между изменением силы, развиваемой мышцей, и изменением числа MHs, которые ее генерируют, совершенно очевидна. Дискутировать здесь можно лишь о том, насколько велика роль этого эффекта (Piazzesi et al. 2007). При этом, однако, нам кажется принципиально важным выяснить, какой именно механизм регулирует число MHs, участвующих в данный момент в генерировании силы. Широко распространено представление о существовании "cooperativity of myosin molecules through strain-dependent chemistry", see e.g. (Duke 2000; Nyitrai and Geeves 2004). Вполне понятно, что strain-sensitive ADP-release mechanism может регулировать длительность isometric contraction state. Тем не менее, совершенно непонятно, о каком именно strain идет речь при объяснении кооперативных явлений. Если это натяжение шейки, которая крепит отдельную MH to thick filament, то при чем тут cooperativity? Если же имеется в виду натяжение филаментов, то каким образом может каждая отдельная MH «узнать» о том, как сильно натянут filament, тем более, что от точки к точке вдоль него натяжение меняется от нуля до максимума?

8Преимущество предлагаемой здесь простейшей модели в том, что прерывание working stroke когда сдвиг МН достигнет максимальной величины $S$ выводит на первый план действительно общую для всех MHs механическую координату – скорость скольжения филаментов $v$. Таким образом, cooperativity оказывается связанной с кинематической характеристикой системы, что вполне соответствует представлениям Huxley A F (2000, p. 435): ''The fact … implies that the molecular events are directly affected by the longitudinal displacement of filaments rather than by the tension in them''.

Более того, в этой модели без каких-либо дополнительных предположений мы получаем такой же знаменатель $\left(1+ v/v_0\right)^{-1}$, что и в Hill's hyperbola (Hill 1938). Поэтому добавляя теперь к (9) hindering force (1) и учитывая, что число detached MHs $N_d = N - N_a$ также зависит от $v$, мы получаем Hill's hyperbola в явном виде:

$$F(v) = F_{prop}(v) - F_h(v) = N f_0 \frac{1 - v/v_{max}}{1 + a \, v/v_{max}}, \quad a = \frac{f_0}{f_h}, \quad v_{max} = a \frac{t_r}{S}. \tag{10}$$

Смысл этой формулы совершенно ясен. С ростом скорости падает число генерирующих силу MHs и растёт число detached MHs, а с ним и число тормозящих движение weak bonds. При максимальной скорости сокращения $v_{max}$ вся энергия, полученная при гидролизе ATP, переходит в тепло т.к. propelling and hindering force становятся равными

$$N_a(v_{max}) f_0 = N_d(v_{max}) f_h \;\Rightarrow\; f_0 = f_h \, v_{max}/v_0. \tag{11}$$

Поэтому параметр $a$, регулирующий форму гиперболы Хилла, оказывается равным просто отношению удельных сил тяги и трения. При этом сама величина максимальной скорости регулируется единственным свободным параметром - продолжительностью recovery time $t_r$.

Используя для hindering force формулу (3) вместо (1), мы получим

$$F(v) = F_{prop}(v) - F_h(v) = N f_0 \frac{1 - \left(v/v_{max}\right)^2}{1 + a \, v/v_{max}}, \quad a = \frac{f_0}{f_h(v_{max})}, \quad v_{max} = \sqrt{\frac{2 E_{ATP}}{p \eta k \tau t_r}}. \tag{12}$$

Теперь скорость $v_{max}$ зависит от большего числа параметров, хотя смысл её остался тем же: при этой скорости $F_{prop}$ and $F_h$ становятся равными. Кроме того, сама force-velocity curve уже отличается от Hill's curve, т.к. в числителе теперь стоит квадрат скорости. Тем не менее, подбирая величину $a$, удаётся сделать кривые (10) и (12) довольно близкими, см.

Fig.2. Более того, кривая (12) при малых скоростях составляет с осью абсцисс на Fig. 2 значительно больший угол, чем при малых силах – с осью ординат. Это похоже на результаты by Ritchie and Wilkie (1958) and Edman (1979), см. также (Rosenfeld 2012, p. 743).

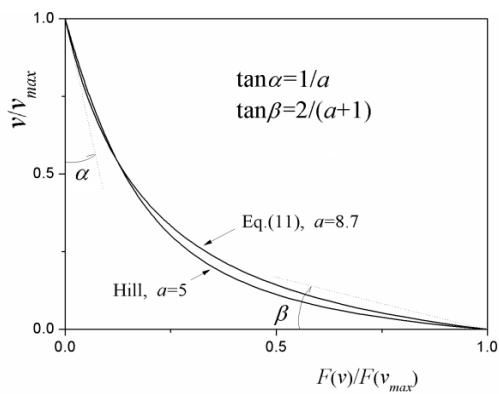

**Figure 2**. Hill's (1938) гипербола (10) с $a$=5 и кривая (12) с $a$=8.7 практически совпадают при больших скоростях. Угол $\alpha$, который кривая Eq. (12) составляет с осью ординат, примерно вдвое меньше угла $\beta$, который она составляет с осью абсцисс.

При вычитании из (9) hindering force (4) в числителе возникают и линейное и квадратичное по $v$ слагаемые, но последнее оказывается положительным. В этом случае, наоборот, угол $\alpha$ на Fig. 2 больше угла $\beta$, что не соответствует эксперименту, и это заставляет думать, что случаи (10) или (12) могут оказаться ближе к реальности. Впрочем, однозначно решить вопросы о зависимости hindering force от скорости, о времени жизни связи $\tau$ и т.д. может только эксперимент. Отметим, наконец, при анализе различных вариантов формул типа (10), (12) следует учитывать, что соотношение $t_w = S/v$ вряд ли может быть справедливым при малых скоростях $v$, see (Rosenfeld 2012).

## V. Conclusions

1. Диссипация той части энергии, выделившейся при гидролизе ATP, которая не превратилась в механическую работу, может происходить в процессе разрыва weak bonds (nonstereospecifical attachments) of MHs to AF.
2. Для проверки этой гипотезы необходимы аналогичные (Bormuth et al. 2009) экспериментальные измерения силы трения между actin and myosin filaments. Вероятно, поставить такой эксперимент можно только если существует аналог ATP, заставляющий MHs по окончании working stroke отделиться от AF, но не дающий им возможности прикрепиться обратно stereospecifically и начать новый цикл.

3. Наличие силы трения, действующей вне MHs, позволяет объяснить значительные различия в скорости релаксации мышечного волокна после резких сокращений и растяжений.

4. Наличие силы трения, величина которой пропорциональна числу detached MHs, вместе с предположением о прерывании working stroke когда сдвиг MH достигнет максимальной величины, позволяет объяснить Hill's force-velocity curve.